%% file: main.tex
\documentclass[conference]{IEEEtran}
\IEEEoverridecommandlockouts
\usepackage{cite}
\usepackage{amsmath,amssymb,amsfonts}
\usepackage{algorithmic}
\usepackage{graphicx}
\usepackage{textcomp}
\usepackage{xcolor}
\usepackage{multirow}
\usepackage{float}
\usepackage{tikz}
\usepackage{lipsum}
\usepackage[caption = false]{subfig}
\def\BibTeX{{\rm B\kern-.05em{\sc i\kern-.025em b}\kern-.08em
    T\kern-.1667em\lower.7ex\hbox{E}\kern-.125emX}}

\newcommand\blfootnote[1]{%
  \begingroup
  \renewcommand\thefootnote{}\footnote{#1}%
  \addtocounter{footnote}{-1}%
  \endgroup
}

\begin{document}

\title{Networks of Ethereum Non-Fungible Tokens:\\ A graph-based analysis of the ERC-721 ecosystem}

\input{authors}

\maketitle

\blfootnote{\scriptsize\copyright 2021 IEEE - Personal use of this material is permitted. Permission from IEEE must be obtained for all other uses, in any current or future media, including reprinting/republishing this material for advertising or promotional purposes, creating new collective works, for resale or redistribution to servers or lists, or reuse of any copyrighted component of this work in other works.}


\input{tex/abstract}
\input{tex/introduction}
\input{tex/soa}

\input{tex/erc721}

\input{tex/data_collection}

\input{tex/general_graph_prop}
\input{tex/time_behaviour}

\input{tex/conclusions}

\bibliographystyle{IEEEtran}
\bibliography{main}

\end{document}

%% file: authors.tex
\author{
\IEEEauthorblockN{S. Casale-Brunet$^1$, P. Ribeca$^2$, P. Doyle$^3$, M. Mattavelli$^1$}
\IEEEauthorblockA{EPFL SCI STI MM, École Polytechnique Fédérale de Lausanne, Switzerland$^1$\\
Biomathematics and Statistics Scotland, United Kingdom$^2$\\
DegenData.io, PinkSwanTrading Inc., United States of America$^3$
}
}
\vspace{-60pt}

%% file: tex/abstract.tex
\begin{abstract}


Non-fungible tokens (NFTs) as a decentralized proof of ownership represent one of the main reasons why Ethereum is a disruptive technology. This paper presents the first systematic study of the interactions occurring in a number of NFT ecosystems. We illustrate how to retrieve transaction data available on the blockchain and structure it as a graph-based model. Thanks to this methodology, we are able to study for the first time the topological structure of NFT networks and show that their properties (degree distribution and others) are similar to those of interaction graphs in social networks. Time-dependent analysis metrics, useful to characterize market influencers and interactions between different wallets, are also introduced. Based on those, we identify across a number of NFT networks the widespread presence of both investors accumulating NFTs and individuals who make large profits.


\end{abstract}


%% file: tex/introduction.tex
\section{Introduction}
\label{s:introduction}


People often surround themselves with objects reminding them of important moments in their lives and experiences. As a result of the digital revolution, what is considered a memory has changed dramatically over the last two decades. For example, photographs that were once printed and stored in albums or other physical media are now always available on electronic devices, thanks to the advent of cloud services and smartphones continuously connected to the Internet. The extensive use of distributed storage raises numerous questions about the ownership and conditions of use of digital objects. The major problems arise because currently, according to the rules governing intellectual property licensing and online contracts originally designed for the Internet of the 2000s, buyers of digital goods are simply users, not owners~\cite{tokenizedFairfield}.

The disruptive element introduced by non-fungible tokens (NFTs) is that NFTs are personal property, not contracts or pure intellectual property licenses~\cite{tokenizedFairfield,jpab104}.  
NFTs allow to demonstrate the ownership of a digital token in a decentralized way (i.e., with no need for a central institution issuing and recognizing a certificate) so that the token can be used in various social spaces, today still typically linked to the digital context. Digital art was one of the first applications of NFTs technology. The likely reason is because of the ease and ability to showcase the item owned, and the possibility to profit from the increased value of a unique piece upon resale. This fact has led an increasing number of artists, including famous ones, to become interested in this new concept.
However, the real media limelight of NFTs has come through the use of this technology in collectibles: as of this writing, in July 2021, the overall market capitalization of the top 100 major NFTs collections is around \$14,492,390,000~\cite{coingecko_nft}. In 2017 CryptoPunk and Crypto-Kitties made history in NFT technology -- the former because some of them were sold for the equivalent of millions of dollars, the latter for making the Ethereum network unusable for a few days due to the high number of transactions originated by their exchange. Later NFT collections from 2021 are, for example, the HashMasks, Meebits, and Bored Ape Yacht Clubs~\cite{opensea_stat}. They are all composed of a limited set of assets characterised by different levels of rarity. Each collection is supported by a cohesive and generally large community communicating on social media platforms such as Twitter or Discord. Another possible use case for NFTs is property. This concept is not yet fully implemented in the real world but very successful in the metaverse, where video game users can buy and trade parts of collectible virtual universes. 

Even if NFTs might become in the next years a powerful example of digital personal property recognized through a decentralized entity, the technology is still in its infancy. Newcomers may get lost in its frenetic evolution due to the lack of systematic summaries and the abundance of hype cycles. This sometimes obscures the properties of a very interesting ecosystem whereby items can be exchanged for very high prices, sometimes hundreds or thousands of ETH. In addition, natural parallels can be drawn between NFT transaction networks and graphs describing the interactions between social media users. The study of the latter has led to algorithms capable of finding authoritative sources on the web or determining user preferences; similarly, exploring the former might lead to analysis methods able to identify trusted or influential users (wallets) and anticipate market evolution. 

To the best of the authors' knowledge, this work is the first to propose a systematic analysis of the dynamics governing the evolution of NFT-related communities in terms of their interaction graphs and associated properties. The main novelties and findings that this paper presents are: (1) A systematic methodology for the analysis of NFT communities; (2) The analysis of such communities over time;
(3) The identification of hub nodes (i.e., wallets) and super-nodes (i.e., wallets across different NFT collections) that influence the market.

The paper is structured as follows:
Section~\ref{s:soa} presents the state of the art in the analysis of graphs related to web networks, social networks and Ethereum ERC20 (fungible) tokens. Section~\ref{s:erc271} summarizes what an Ethereum ERC-721 smart contract is. Section~\ref{s:dataCollection} describes which NFT collections have been analysed for this work and how the data has been collated; crucially, it defines the structure of an NFT transaction graph. Section~\ref{s:generalGraphProperties} illustrates the topological and clustering analysis of such transaction graphs. Section~\ref{s:timeBehaviour}  presents the methodologies for estimating the value of a collection and how to identify from the flows between major wallets and most successful investors. Finally, section~\ref{s:conclusion} concludes the paper and provides some ideas for possible further analysis.

%% file: tex/soa.tex
\section{Related work}
\label{s:soa}


Some of the metrics most commonly used to identify trends in the cryptocurrency market include technical analysis (i.e., studying price charts to identify opportunities), fundamental analysis (i.e., studying the economic and technological factors underlying a project), and metrics extracted from social media. The latter can help an investor take the pulse of what the general public is focusing on; analyzing the mentions a project gets on Twitter or examining Google Trends data are other examples of alternative ways traders monitor emerging trends and industry cycles in the cryptocurrency ecosystem. 

Analysis of web searches and how more relevance is given to one web page than another is a field with a flourishing literature. In general, each web page is modelled as a node and the connections between them as graph links~\cite{web_graph_andrei}. Analyses of this kind are for instance at the heart of search engines (e.g., Google, Baidu, Microsoft Bing). They are also used to understand how to produce increasingly effective web content and advertising campaigns, and to identify trends that are emerging among users~\cite{kil2009graph,meusel2014graph}. 

Similarly, data analysis of social networks generally involves modelling the network(s) as a directed graph where the nodes are the users and the connections are the iterations between them.  Seminal work on this type of analysis dates back to 2007, i.e. just after the foundation of Facebook (2004) and Twitter (2006), with the work presented in~\cite{social_network_graph_mislove}. In this work, which was the first to propose a methodology to examine multiple online social networks at scale and related examples, roughly 11.3 million users and 328 million links were analyzed. By measuring the topological properties of these graphs, the groundwork was established to define key concepts, for instance that influential and hub nodes can drive the sentiment of a large number of network nodes~\cite{campbell2013social,social_network_analysis_wright,papadopoulos2012community}. 

Similar analysis techniques have been applied to blockchain transactions. Specifically, much of the research has focused on analyzing transactions in the Bitcoin blockchain to define flows and de-anonymize wallets. However, graph analysis seems much better suited to account-based blockchains like Ethereum rather than UTXO-based like Bitcoin~\cite{social_network_graph_mislove,eth_graph_erc20,eth_graph_chen,eth_temporal_zhao}. With respect to Ethereum, some current research has focused on the analysis of ERC20~\cite{EIP20} (fungible) token transactions. In~\cite{eth_graph_erc20} it has been shown how the entire (global) network of ERC20 token transfers follows a power-law in its degree distribution, but this cannot be stated for many individual token networks. This is because many of the ERC20 token networks follow either a star or a hub-and-spoke pattern. The heavy tails in the degree distribution are not as pronounced as in social networks: ERC20 token networks tend to contain less and smaller hubs. Furthermore, the main use case of many such tokens appears to be for trading and the number of exchanges is limited, meaning that only a few agents succeed in gaining the trust of the users. This leads to big hubs that can potentially obscure the real impact on market dynamics had by other smaller but more influential nodes. 

%% file: tex/erc721.tex
\section{ERC-721 non-fungible token standard}
\label{s:erc271}

A NFT is used to uniquely identify something or someone. For instance, typical use cases are related to collectibles, access keys, tickets, numbered seats for concerts and sport events. ERC-721 (Ethereum Request for Comments 721, see~\cite{EIP721}) standardizes non-fungible tokens by defining an API interface to implement a smart contract on the Ethereum blockchain and by mandating the functionality that such a smart contract must provide. 
The functionality is mainly related to the transfer of tokens from one account to another, the retrieval of the current balance of a wallet, and the retrieval of the ownership of a specific token. As an example of a ERC-721 smart contract, the interested reader can refer to the deployed source code of the HashMasks smart contract~\cite{hashmasks_smartcontract}.
 
\input{tables/nft_names}

%% file: tables/nft_names.tex
\begin{table}[htbp]
\begin{center}
\caption{NFTs collection summary in terms of category and Ethereum ERC-721 smart contract address.}
\label{t:nft:names}
\resizebox{0.4\textwidth}{!}{
\begin{tabular}{|l|l|l|}
\hline
\textbf{Collection}            & \textbf{Category} & \textbf{Smart Contract Address} \\\hline
HashMasks             & Digital art     &  {\tiny 0xc2c747e0f7004f9e8817db2ca4997657a7746928} \\\hline
Art Blocks Curated    & Digital art     &  {\tiny 0xa7d8d9ef8d8ce8992df33d8b8cf4aebabd5bd270} \\\hline   
Cryptopunks           & Profile picture &  {\tiny 0xb47e3cd837ddf8e4c57f05d70ab865de6e193bbb} \\\hline   
Bored Ape Yacht Club  & Profile picture &  {\tiny 0xbc4ca0eda7647a8ab7c2061c2e118a18a936f13d}\\\hline   
Acclimated Moon Cats  & Profile picture &  {\tiny 0xc3f733ca98e0dad0386979eb96fb1722a1a05e69} \\\hline   
CryptoVoxels          & Metaverse       &  {\tiny 0x79986af15539de2db9a5086382daeda917a9cf0c} \\\hline   
Decentraland          & Metaverse       &  {\tiny 0xf87e31492faf9a91b02ee0deaad50d51d56d5d4d} \\\hline   
Meebits               & Metaverse       &  {\tiny 0x7bd29408f11d2bfc23c34f18275bbf23bb716bc7} \\\hline   
\end{tabular}
}
\end{center}
\end{table}

%% file: tex/data_collection.tex
\section{Collecting and representing data}
\label{s:dataCollection}

\input{tables/nft_description}
\input{tables/nft_collected_data}

All the analyses described below were performed on a desktop computer equipped with a quad-core Intel i7 Kaby Lake CPU and 16 GB of RAM, running MacOS Big Sur 11.4. Each NFT transaction dataset ranged from 10 to 20 MB in size.

\subsection{Selected NFT projects}
In this work, 8 different NFT projects have been analyzed. These are reported in Table~\ref{t:nft:names} along with the Ethereum address of their smart contracts. The choice of these 8 different projects is motivated by the fact that these are, to the best of the authors' knowledge, the most significant ones in terms of both market capitalisation and popularity, across categories such as profile pictures, curated digital art and metaverse. Table~\ref{t:nft:summary} reports the projects summary in terms of unique assets (i.e., available tokens), net worth, number of different owners (i.e., wallets), traded volume and floor price reported the 15th July 2021. The net worth value has been retrieved from~\cite{coinmarketcap_nft}, the CryptoPunks floor price from~\cite{cryptopunks_floor}, while the other values come from~\cite{opensea_stat}. Furthermore, as with the rest of the article, the value of 1 ETH has been normalized to 2000\$ from the daily price retrieved from~\cite{coinmarketcap_eth_price} so as to make values comparable. Unfortunately, as illustrated by the table, some values are unavailable even after aggregating data from different websites. Despite the fact that market capitalization is growing rapidly, the main reason why this crucial information is missing is a generalized lack of services offering complete and detailed analytics for all the existing NFT projects.

\subsection{Data extraction}
Blockchains store massive amounts of heterogeneous data that grows over time. Focusing on the Ethereum network, when searching for information on the blockchain it is necessary to directly access records (blocks) using a unique identifier (e.g. block ID, transaction, wallet or contract address)~\cite{wood2014ethereum}. As a result, effective techniques for data storage and extraction must be adopted~\cite{eth_data_extraction_brinckman} in order to conduct analysis effectively without having to sequentially walk through all the blockchain data. The last few years have seen the creation of services, some of them free and other ones commercial, capable of storing blockchain information daily and incrementally, in specific infrastructures suited to the analysis of Big Data. These can then be subsequently queried using traditional SQL syntax (see~\cite{eql_bragagnolo,eth_sql_han}). For the purpose of this work, the Google BigQuery platform~\cite{google_bigquery} has been used. Blockchain data was extracted in an organized way that would facilitate subsequent analysis. The results were stored in comma separated value (CSV) files, 
accessible to the interested reader from~\cite{dataset_lin}.
The fields include information such as the hash of the transaction, the address of the NFT smart contract, the addresses of the wallet(s) or smart contract(s) that sold and purchased the NFT token, the Ethereum block time as milliseconds from the UNIX epoch, the Ethereum block number, the value of ETH and/or WETH (Wrapped ETH) transferred and the number of tokens exchanged during the transaction (can be more than one). It should be noted that a few transactions involve ETH/WETH transfers to more than one address or transfers back to the originating address, resulting in more than one record having the same transaction hash. The simplifications made for the queries were: when the seller goes through a proxy contract (e.g., see~\cite{transaction_example_maxeth}) the maximum amount of ETH and/or WETH traded is considered as a cost; gas fees (i.e., transaction costs) are excluded from the analysis; if the exchange currency is neither ETH nor WETH the transaction is seen as a simple transfer (seen that the vast majority of the transactions happen in ETH or WETH, we evaluated that this assumption still produces reasonable estimates of the exchange volumes while significantly reducing the complexity of the queries used). Whenever more than one token is exchanged in the same transaction, the individual price of each token is considered the same, and computed by splitting the total amount of the transaction into equal parts. For each NFT collection, data was harvested from the date of their creation to July 15th, 2021.

\subsection{Bulk statistics}
As a first analysis, we defined and summarized two types of account addresses and three different types of transactions. As per general Ethereum terminology~\cite{ethereum_whitepaper}, there are two different types of accounts:
\begin{itemize}
    \item EoAs: externally owned accounts, controlled by private keys (i.e., a general purpose personal wallet);
    \item CAs: contract accounts, controlled by their contract code. 
\end{itemize}
We also categorized transactions into three different types:
\begin{itemize}
    \item Buy/Sell: Indicates a transaction where the selling address is different from the buying address and where the buying address has exchanged the token for a non-zero amount of ETH and/or WETH;
    \item Transfer: The selling address is the same as the buying address, and/or the transaction has been made with a non-zero amount of ETH and WETH; 
    \item Mint: The selling address is \texttt{\small 0x000000}. Minting transactions are special ones that are generally used to create an initial distribution of tokens. This is done, for example, by invoking a minting function in the smart contract until all assets are distributed (see for instance the \texttt{\small mintNFT()} function in the HashMasks contract~\cite{hashmasks_mint}).
\end{itemize}
Statistics for the first category are summarized in Table~\ref{t:nft:collected_data}. For self-transactions (see e.g.~\cite{transaction_example_self}) the exchanged ETH and/or WETH value does not contribute to the traded volume.

\subsection{Transactions graph model}
\label{s:graphModel}
The collected transactions can be structured in a multi-directed weighted graph $MDG(V,E)$ with a set of nodes $V$ and edges $E = \{ (v_i, v_j)^k ~|~ v_i \in V, v_j \in V, k \in \mathbb{N}^+ \}$. Each node $v_i \in V$ represents a unique wallet, and each directed edge $e_{i,j}^k \in E$ represents a single token transaction from $v_i$ to $v_j$. For each transaction, additional information is defined and stored as an edge parameter. Examples are the weight $w(v_i, v_j)^k$ corresponding to the cost payed by the wallet $v_j$ to the wallet $v_i$ to perform the transaction (i.e., to buy the token); the ID of the transferred token, $T(v_i, v_j)^k$; and the transaction time $\tau(v_i, v_j)^k$.

%% file: tables/nft_description.tex
\begin{table*}[htbp]
\begin{center}
\caption{NFTs collection summary in terms of unique assets, net worth, number of different owners, traded volume and floor price reported the 15th July 2021. The net worth value has been retrieved from~\cite{coinmarketcap_nft}, the CryptoPunks floor price from~\cite{cryptopunks_floor}, while the other values from~\cite{opensea_stat}. 1 ETH has been considered as 2000\$ as the daily price retrieved from \cite{coinmarketcap_eth_price}.}
\label{t:nft:summary}
\resizebox{0.6\textwidth}{!}{
\begin{tabular}{|l|c|r|r|c|r|r|r|r|r|}
\hline
\multirow{2}{*}{\textbf{Collection}} &  \multirow{2}{*}{\textbf{Asset}} & \multicolumn{2}{c|}{\textbf{Net Worth}}  & \multirow{2}{*}{\textbf{Owners}} & \multicolumn{2}{c|}{\textbf{Volume Traded}} & \multicolumn{2}{c|}{\textbf{Floor Price}} \\
& & \textbf{ETH} & \textbf{USD} & & \textbf{ETH} & \textbf{USD} & \textbf{ETH} & \textbf{USD} \\\hline
HashMasks             & 16384  & 32826.6  & 65653218  & 4500 & 30900  & 61800000  & 0.42  &  840   \\\hline
Art Blocks Curated    & 57873  & 24819.5  & 49639177  & 4100 & 16800  & 33600000  & 0.20  &  400   \\\hline
CryptoPunks           & 10000  & 195188.4 & 390376817 & 2500 & 196800 & 393600000 & 19.75 &  39500 \\\hline
Bored Ape Yacht Club  & 10000  & -        & -         & 4700 & 30500  & 61000000  & 4.70  &  9400  \\\hline
Acclimated Moon Cats  & 10765  & -        & -         & 1700 & 660    & 1320000   & 0.25  &  500   \\\hline
CryptoVoxels          & 14872  & 6136.7   & 12273569  & 1300 & 16500  & 33000000  & 1.06  &  2120  \\\hline
Decentraland          & 177094 & 33223.8  & 66447728  & 4700 & 90800  & 181600000 & 1.62  &  3240  \\\hline
Meebits               & 20000  & 77381.8  & 154763654 & 4700 & 25300  & 50600000  & 1.88  &  3760  \\\hline
TOTAL (sum)           & 315491 & -        & -         & -    & 408260 & 816520000 & -     &  -     \\\hline
\end{tabular}
}
\end{center}
\end{table*}

%% file: tables/nft_collected_data.tex
\begin{table*}[htbp]
\begin{center}
\caption{NFTs collected data summary in terms of wallets, transactions and volume.}
\label{t:nft:collected_data}
\resizebox{0.65\textwidth}{!}{
\begin{tabular}{|l|r|r|r|r|r|r|r|r|r|r|} \hline
\multirow{2}{*}{\textbf{Collection}} & \multicolumn{3}{c|}{\textbf{Wallets}}        & \multicolumn{3}{c|}{\textbf{Transactions}}                            & \multicolumn{4}{c|}{\textbf{Volume (ETH)}}                 \\
                                     & \textbf{Tot} & \textbf{EoAs} & \textbf{CAs} & \textbf{Tot} & \textbf{Buy/Sell} & \textbf{Transfer} &  \textbf{Tot} & \textbf{Max} & \textbf{Avg} & \textbf{Var} \\ \hline
HashMasks	& 7730	& 7250& 	480& 	49404& 	15624& 	17396& 		37331.35& 	420.00& 	2.38 & 	43.17 \\\hline
Art Blocks Curated& 	9591& 	9541& 	50& 	80660& 	20144& 	11779& 		17777.18 & 	65.00 & 	0.88 & 	4.69  \\\hline
Cryptopunks	& 4631& 	4139& 	492	& 28576	& 10459	& 18117	&  154195.37	& 4200.00& 	14.74 & 	2311.11 \\\hline
Bored Ape Yacht Club & 	7214& 	7185& 	29	& 32074	& 15874	& 6200& 	 	31092.40 & 	109.00	& 1.95& 	8.31 \\\hline
Acclimated Moon Cats& 	2045& 	2036& 	9& 	14378& 	1739& 	1874& 	 	800.92 & 	29.00 & 	0.46 & 	1.36  \\\hline
CryptoVoxels& 	2085& 	2048& 	37& 	16607& 	8394& 	3090& 	 	33217.11 & 	250.00 & 	3.95 & 	317.63 \\\hline
Decentraland& 	5045& 	4949& 	96& 	174322& 	540	& 116151& 		3321.17	& 78.92 & 	6.15 & 	133.39 \\\hline
Meebits& 	6448& 	6419& 	29& 	29224& 	6250& 	2974& 		27181.02 & 	1000.00 & 	4.34 & 	503.74 \\\hline
ALL (merged)& 	33675& 	32510& 	1165& 	425245& 	79024& 	177581& 	304916.56 & 	4200.00 & 	3.85 & 	411.16 \\\hline       
\end{tabular}
}
\end{center}
\end{table*}

%% file: tex/general_graph_prop.tex
\begin{figure*}
\subfloat[HashMasks\label{f:graph:hm}]{\includegraphics[width = 0.23\linewidth]{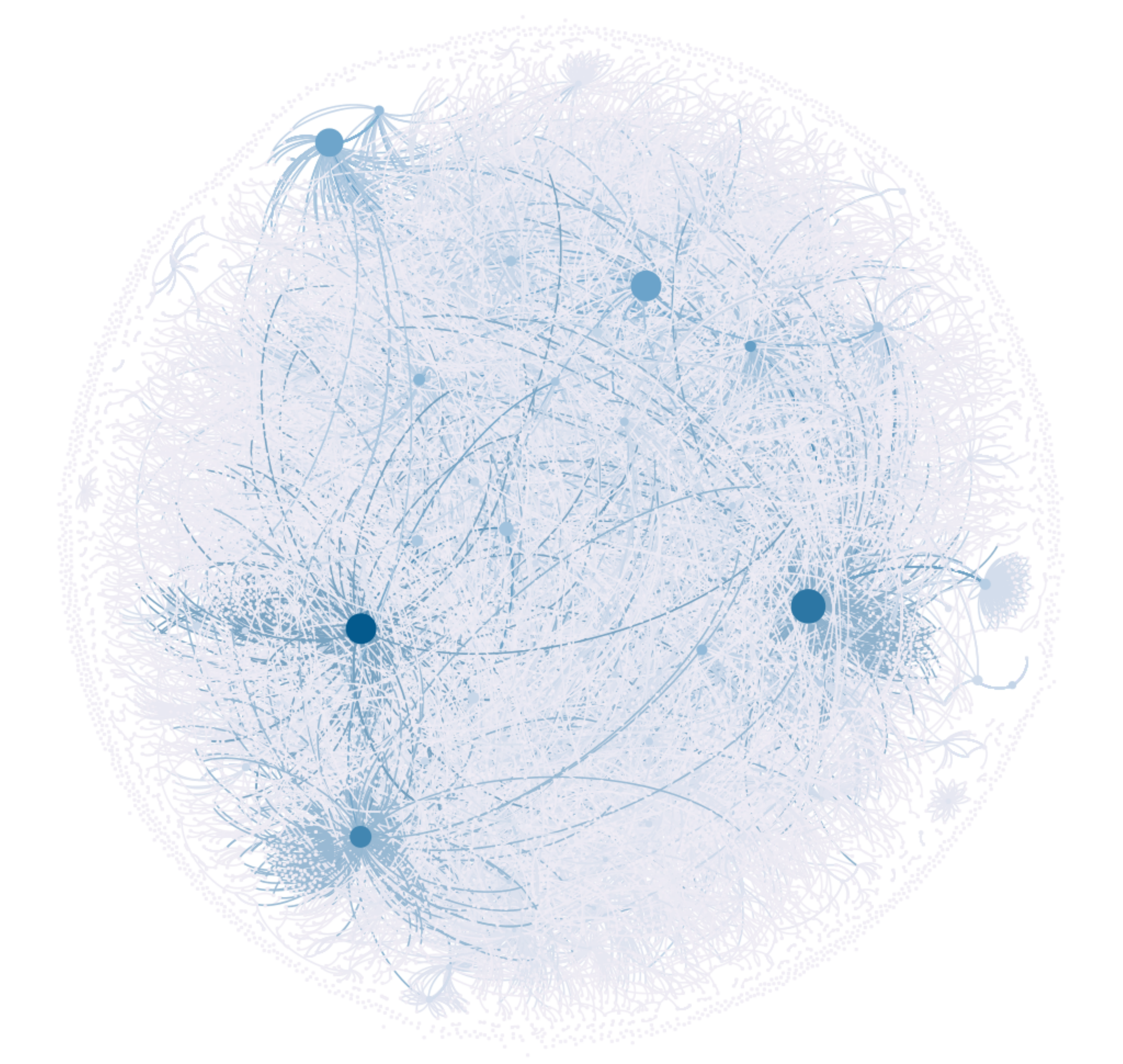}}
\subfloat[Bored Ape Yacht Club\label{f:graph:bayc}]{\includegraphics[width = 0.25\linewidth]{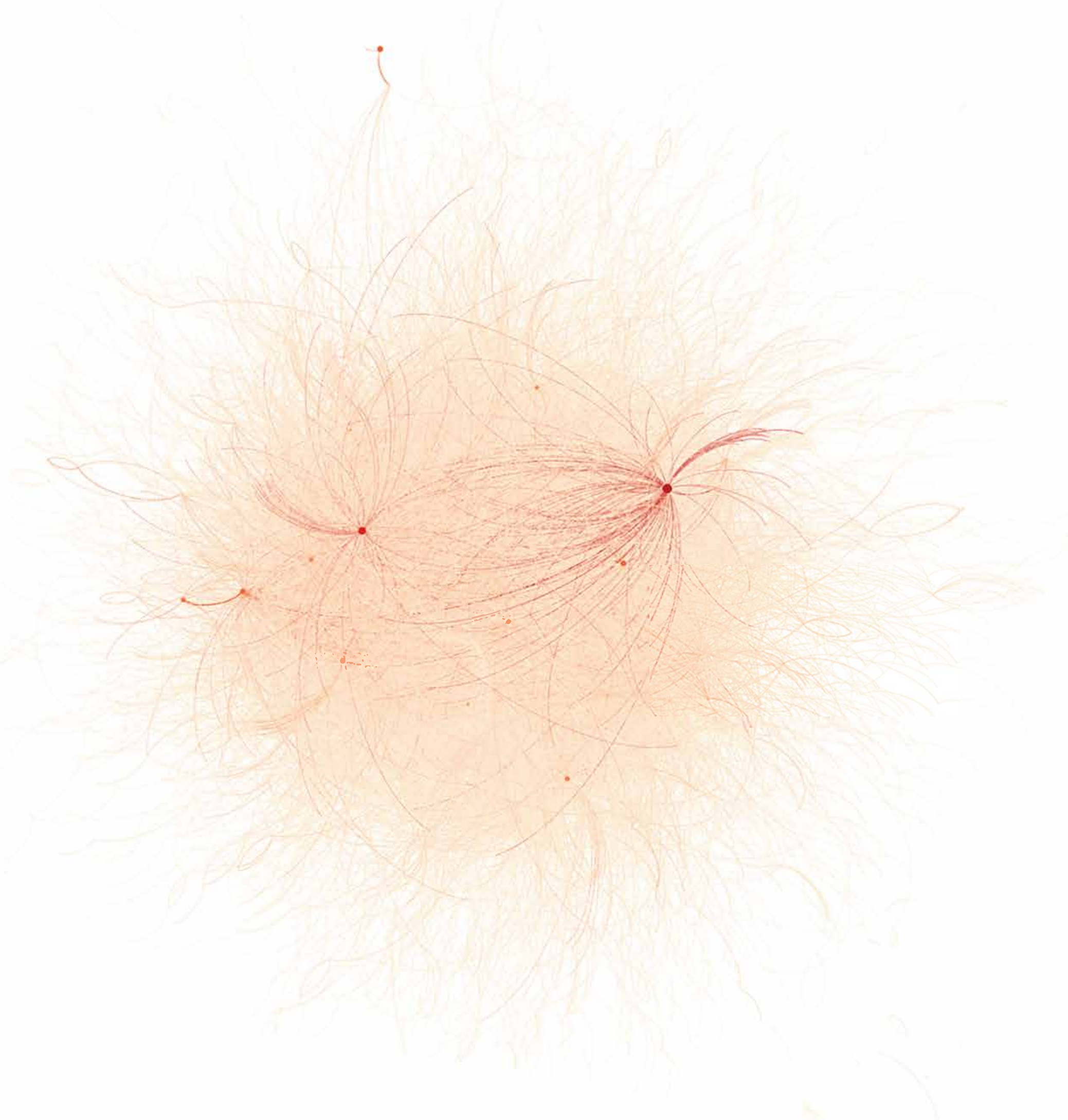}}
\subfloat[Cryptopunks\label{f:graph:cp}]{\includegraphics[width = 0.23\linewidth]{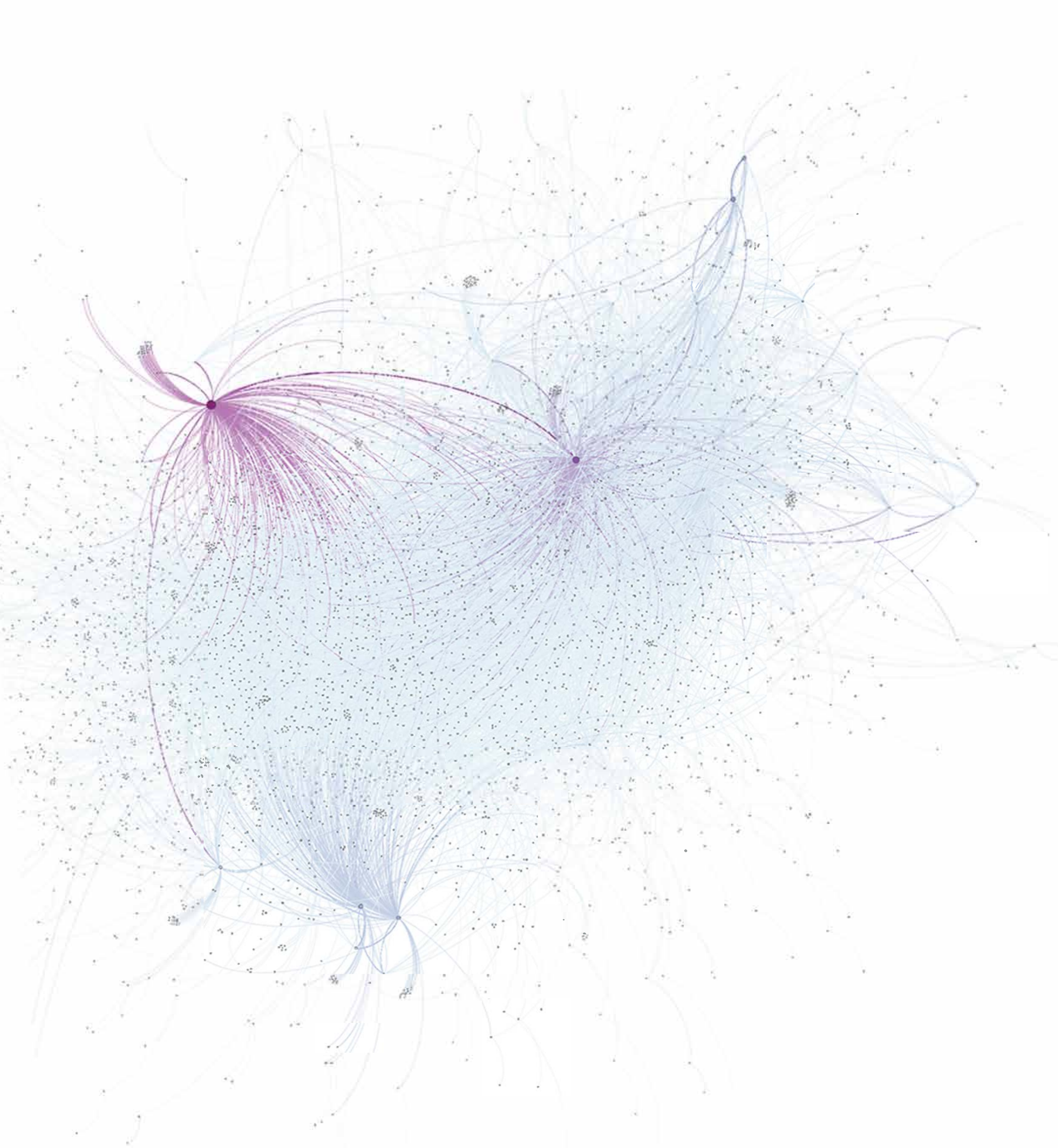}}
\subfloat[All the NFT projects listed in Table~\ref{t:nft:names} \label{f:graph:all}]{\includegraphics[width = 0.3\linewidth]{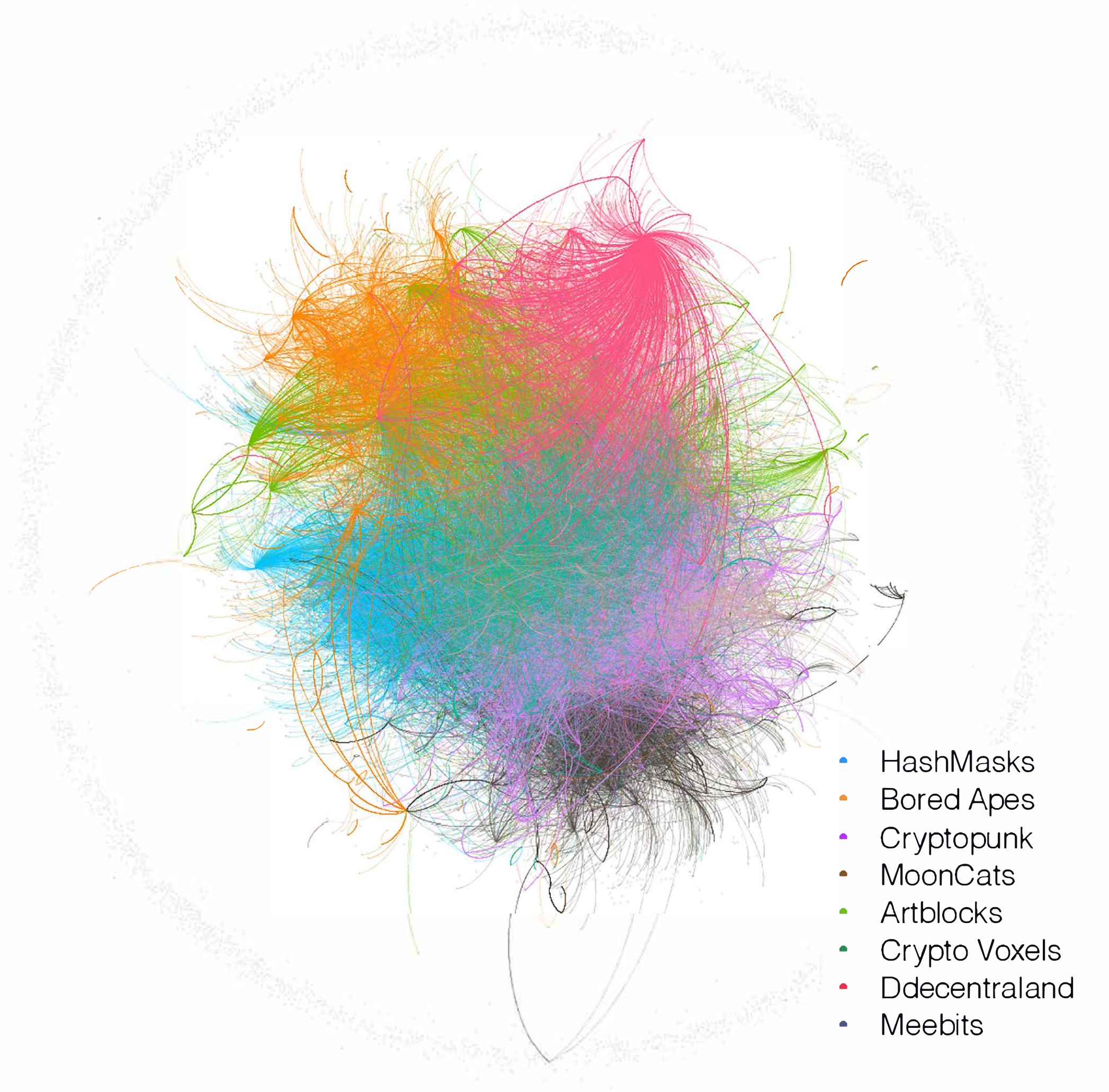}}
\caption{Some of the transaction graphs listed in Table~\ref{t:graph:stats}. The size of the nodes is directly proportional to their input degree (i.e., the larger a node, the more tokens it has accumulated from a given collection). In~(\ref{f:graph:all}) edges are coloured according to the project the transferred token belongs to.}
\label{f:graph:graphs}
\end{figure*}

\section{Transaction graph analysis}
\label{s:generalGraphProperties}
In this section we present the main static properties of the NFT transaction graphs derived from the data collected in Section~\ref{s:dataCollection}. In particular, properties such as distributions of degree, density, component clustering coefficients, and assortativity, are all evaluated and discussed.

The power-law results described in Section~\ref{s:graph:topo} have been obtained using the \textit{poweRlaw} package in R~\cite{poweRlaw_R}.

\input{tables/graph_stats}
\input{tables/graph_powerlaw}
\input{tables/graph_connectivity}
\input{tables/soa_graphs_stats}

\subsection{Basic quantities}
Only transactions between two EoA addresses have been considered in this work. We did so because the main objective of this work is to discover and analyze the behavior of communities participating in the buying and selling of NFTs. Thus, we attempted to eliminate at least some of the biases by discarding all the transactions made by smart contracts and not by generic wallets (the latter likely corresponding to actual users). In addition, we eliminated the mint node: every transaction between the mint node and a generic node of the graph was saved as a parameter of the node itself. This additional parameter, discussed in the next sections, allows to calculate the distribution of node in-degrees and the flows of tokens within the network without disrupting connectivity and clustering analysis.

Table~\ref{t:graph:stats} summarizes measurements for each transaction graph (in terms of number of nodes and edges, diameter and mean distance) when each NFT project is considered separately. It also shows the same quantities for the graph obtained by considering all the projects listed in Table~\ref{t:nft:names} together: in other words, the graph where each wallet can exchange tokens from different NFT projects. To get a visual idea of the structure of these graphs, see Figure~\ref{f:graph:graphs}. It shows the graphs of Hashmasks (Figure~\ref{f:graph:hm}), Bored Ape Yacht Club (Figure~\ref{f:graph:bayc}), Cryptopunks (Figure~\ref{f:graph:cp}); and the overall graph of all the projects examined in this work (Figure~\ref{f:graph:all}). The size of each node is proportional to its input degree (i.e., the larger a node, the more tokens it has accumulated from a given collection). In Figure~\ref{f:graph:all} the edges are coloured according to the project the transferred token belongs to; one can see that real, separated wallet communities do exist according to each project, but also that wallets trading on more than one project are frequent.
Interestingly, the diameter (i.e. the length of the longest path, measured as the number of edges) and the mean distance (i.e. the average number of edges between any two nodes in the network) are very similar to those identified in~\cite{social_network_graph_mislove} from the analysis of social network graphs (see Table~\ref{t:graph:soa}). That is true for individual and combined transaction graphs.

\subsection{General topological properties}
\label{s:graph:topo}

\begin{figure}[htbp]
\begin{center}
    \subfloat{\includegraphics[width = 0.49\linewidth]{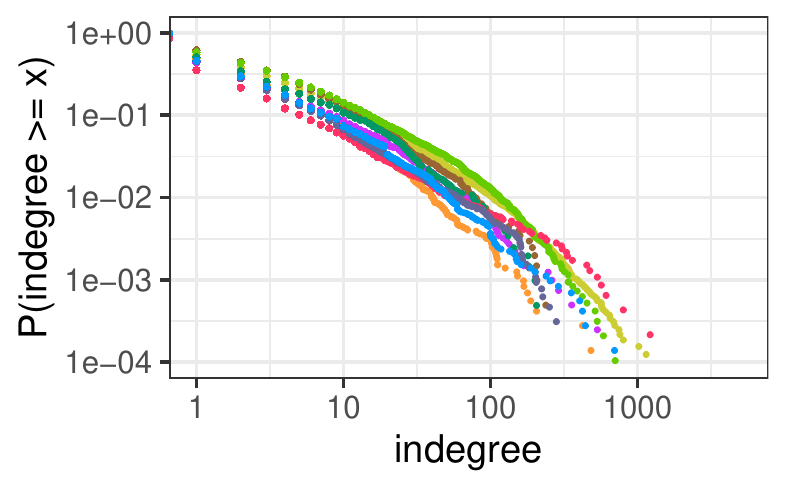}}
    \subfloat{\includegraphics[width = 0.49\linewidth]{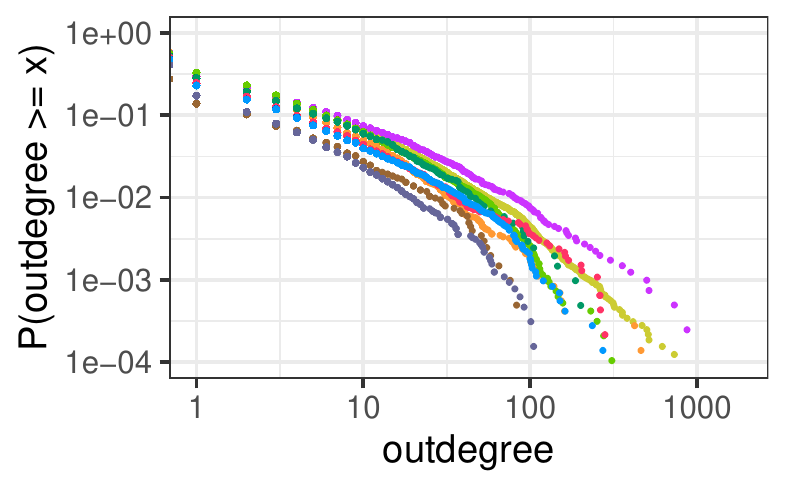}}  \\
    \subfloat{\includegraphics[width = 0.9\linewidth]{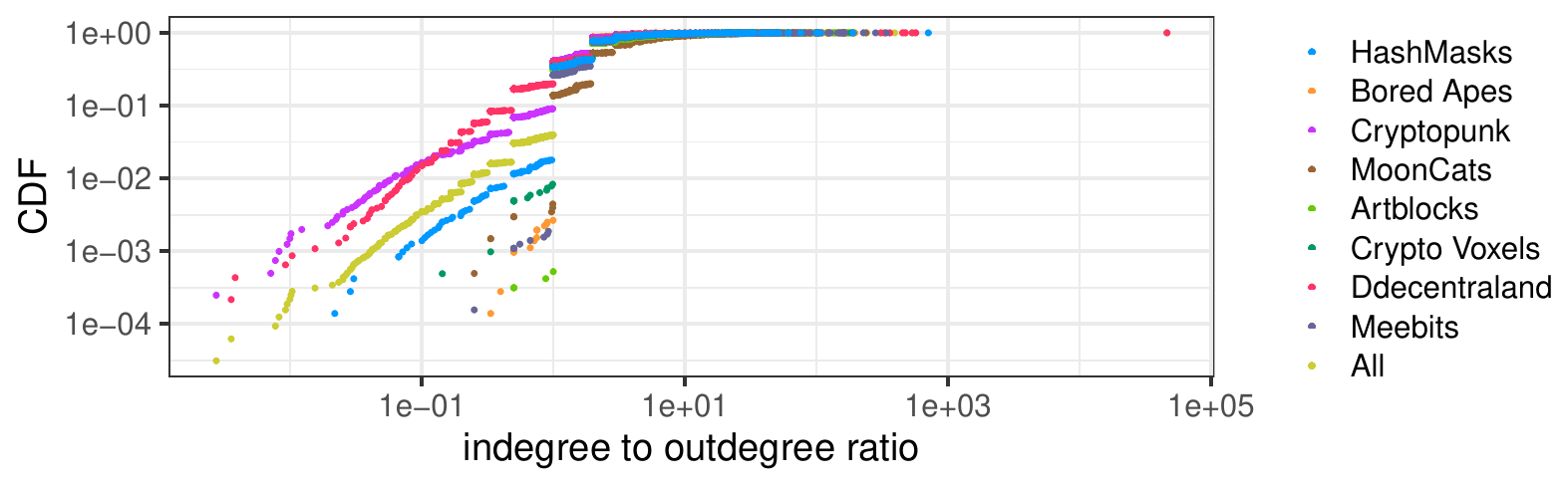}}
\end{center}
\caption{Degree analysis for the different NFT collection graph: complementary in- and out-degree probability distribution, and cumulative distribution of the in- to out-degree ratio. Power-law parameters obtained from the degree probability distribution are reported in Table~\ref{t:graph:powerlaw}.}
\label{f:graph:degree}
\end{figure}

Fundamental properties of a directed graph are in- and out-degree of nodes. 
The frequency distribution of degrees in transaction graphs can provide insight into user behavior when trading a particular collection of NFTs.
Many real-world graphs for social media and the Internet show highly skewed, heavy-tailed degree distributions~\cite{eth_graph_erc20, web_graph_andrei,web_graph_wu,social_network_graph_mislove}. This generally indicates that a significant portion of the information about node interactions needs to be extrapolated from the analysis of their tails, and in general demonstrates the existence of (several) high degree hubs. For example, in Ethereum ERC20 token networks hubs are exchanges~\cite{eth_graph_erc20}, in social networks they are influencer nodes~\cite{social_network_graph_mislove,Vosoughi1146}, and in web networks they are popular and high-ranked websites~\cite{web_graph_wu}.
Several kinds of networks have been confirmed to follow power laws in their degree distribution~\cite{eth_graph_erc20, web_graph_andrei,web_graph_wu,social_network_graph_mislove}. Power laws are distributions of the form $p(x) = Cx^{-\alpha}$, in which the dependent variable, the probability for a node to have degree $k$, varies as an inverse power of the independent variable, the degree $x$. In other words, $p(x)$ decreases monotonically, but significantly slower than the exponential decay of normal distributions. While the non-negative constant $C$ is fixed by normalization, the parameter $\alpha$ is called the coefficient of the power law. Typical values of $\alpha$ are is in the range $2 \leq \alpha \leq 3$. Figure~\ref{f:graph:degree} shows the complementary cumulative distribution function (CCDF) of in- and out-degree, and the cumulative distribution function (CDF) of the in-degree to out-degree ratio for each measured NFT network.

Table~\ref{t:graph:powerlaw} shows the estimated power-law coefficient for each NFT network along with the Kolmogorov-Smirnov goodness-of-fit metric~\cite{clauset2009powerlaw}. The best-fit power-law coefficients approximate the distributions very well for all NFT collections when they are considered separately except for the case of Decentraland. Table~\ref{t:graph:powerlaw} shows the values for graphs built for web network~\cite{web_graph_andrei}, social network~\cite{social_network_graph_mislove}, and the global ERC20 (fungible) token network~\cite{eth_graph_chen}. As for social networks, the values reported in the table represent the mathematical average of the values reported for Flickr, Livejournal, and YouTube (data for Orkut has been excluded because incomplete, as also stated in~\cite{social_network_graph_mislove}). It is worth mentioning that despite many network graphs following a power-law distribution, the ones for a number of ERC20 (fungible) tokens do not do so when considered individually. In fact, in~\cite{eth_graph_erc20} it has been shown that individual networks for ERC20 (fungible) tokens are frequently dominated by a single hub (generally an exchange) and follow either a star or a hub-and-spoke pattern where the heavy tails in the degree distribution are not as pronounced as in social networks. However, this is not the behaviour that we observe for ERC721 (non-fungible) tokens. Interestingly, Figure~\ref{f:graph:degree} panels A and B shows that all the power-law coefficients for the separate networks are similar, but the one for the combined network is different. This is probably due to the different proportion of hubs in the separate and combined network.

Another interesting parameter to analyze is the relationship between in- and out-degree of each node. In other words, this indicates what is the ratio between purchases (i.e., accumulation) and sales for the various wallets that make up the network. As an example, studies on the distribution of in- and out-degree in web networks have consistently helped with the identification of better methods to find relevant information on the web. Generally speaking, in web networks the population of pages that are active (i.e., that have a high out-degree) is not the same as the population of pages that are popular (i.e., that have a high in-degree). As pointed out in~\cite{web_graph_wu}, web-search techniques can be very effective when they are able to separate a very small set of popular pages from a much larger set of active pages. As shown in Figure~\ref{f:graph:degree} panel 3, NFT transaction networks have a large number of active nodes with low in- to out-degree ratio, (i.e., nodes selling NFTs), and a much more limited number of nodes with high ratio (i.e., nodes accumulating NFTs). We can then draw a natural parallel between web influencers and accumulators of NFTs.


\subsection{Connectivity and clustering properties}

\begin{figure}[htbp]
\begin{center}
    \subfloat{\includegraphics[width = 0.5\linewidth]{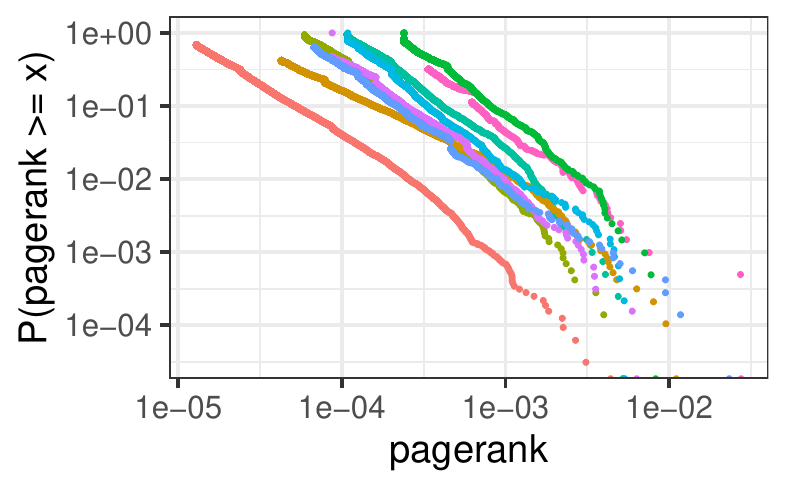}}
    \subfloat{\includegraphics[width = 0.5\linewidth]{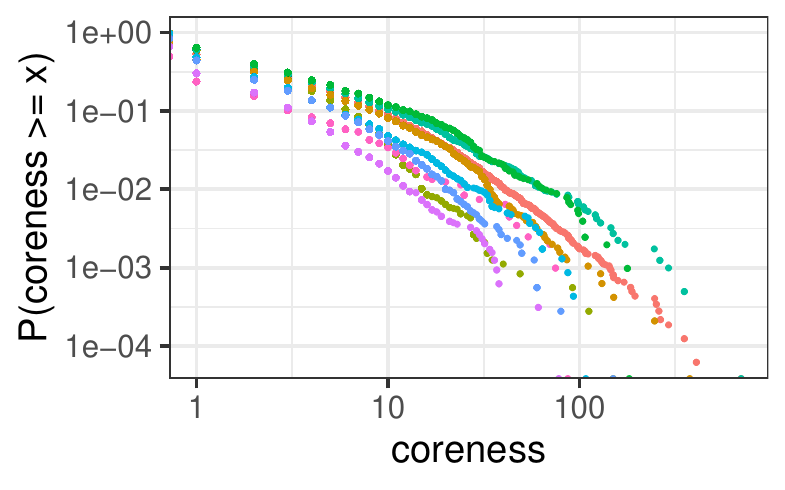}}
\end{center}
\caption{Graph connectivity values in term of CCDF pagerank and coreness distributions. The colours are the same as those of the legend in Figure~\ref{f:graph:degree}.}
\label{f:graph:connectivity}
\end{figure}

The assortativity coefficient~\cite{meghanathan2016assortativity} measures the level of homophyly of the graph (i.e., how nodes are connected with respect to a given property) and its value range is $[-1, 1]$.
%
According to~\cite{meghanathan2016assortativity}, the graph is said to be strongly assortative, weakly assortative, neutral, weakly disassortative, and strongly disassortative, if the assortativity coefficient falls into the ranges $[0.6, 1]$, $[0.2, 0.6)$, $(-0.2, 0.2)$, $(-0.6, -0.2]$, and $[-1, -0.6]$, respectively.
Table~\ref{t:graph:connectivity} illustrates the assortativities for NFT transaction graphs. It shows that most of them are neutral associative, with the exception of Acclimated Moon Cats that are strongly assortative, and ArtBlocks and CryptoPunks that are weakly assortative.
The transitivity coefficient (also known as clustering coefficient) measures the probability for adjacent nodes of a network to be connected -- in other words, if there is a link $e(v_i, v_j)$ or $e(v_j, v_i)$ and a link $e(v_j, v_k)$ or $e(v_k, v_j)$, what is the probability of there being a link $e(v_i, v_k)$ or $e(v_k, v_i)$.
Table~\ref{t:graph:connectivity} illustrates the various levels of transitivity of the graphs: they all tend towards 0, indicating that collaboration between wallets is very rare.
The reciprocity coefficient measures the proportion of mutual connections on a direct graph. In other words, it estimates the probability that given an edge $e(v_i, v_j)$, then its reciprocal $e(v_j, v_i)$ also exists. From Table~\ref{t:graph:connectivity} it is possible to see that reciprocity levels are close to zero, indicating in fact that buy/sell trades are only made in one direction, which is also the case when considering the global graph. In other words, exchanges between wallets are still uncommon.
Figure~\ref{f:graph:connectivity} shows the CCDF of both pagerank and coreness values for the NFT transaction network graphs considered. First, the distribution of pageranks~\cite{brin1998anatomy} in each graph was analyzed. This metric is generally used in web network graphs to estimate the importance of each node. The results show the presence of nodes that are more important than others. On the other hand, coreness is a measure used to identify closely interconnected groups within a network. A k-core is a (maximal) group of entities, all connected to at least $k$ other entities in the same group. As can be seen from the figure, these networks have cores of degrees greater than $100$, albeit with very low probability.
It must be noted that, as demonstrated in~\cite{liu2015core}, high coreness nodes are not always influencers, although it is highly likely that influencer nodes are high coreness nodes.


%% file: tables/graph_stats.tex
\begin{table}[htbp]
\begin{center}
\caption{Network graph statistics for each NFT collection in terms of number of vertices, edges, diameter, and mean distance. The merged graph considers all collections together.}
\label{t:graph:stats}
\resizebox{0.42\textwidth}{!}{
\begin{tabular}{|l|r|r|c|c|} \hline
\textbf{Collection} & \textbf{Vertices} & \textbf{Edges} & \textbf{Diameter} & \textbf{Mean Distance} \\\hline
HashMasks	& 7174& 	16984& 	28& 	4.94\\\hline
Art Blocks Curated& 	9537& 	31144& 	15& 	4.71 \\\hline
Bored Ape Yacht Club & 	7177& 	20244& 	18& 	5.88  \\\hline
CryptoPunks	& 4028& 	20568& 	15& 	5.05 \\\hline
Acclimated Moon Cats& 	2022& 	2751& 	14& 	4.46 \\\hline
CryptoVoxels & 	2040& 	11134& 	15& 	5.24 \\\hline
Decentraland & 	4620& 	12656& 	17& 	4.91 \\\hline
Meebits & 	6413& 	8898& 	18& 	6.51 \\\hline
ALL (merged) & 	32121& 	124379& 	31& 	5.35    \\\hline               
\end{tabular}
}
\end{center}
\end{table}

%% file: tables/graph_powerlaw.tex
\begin{table}[htbp]
\begin{center}
\caption{Network graph power law values for each NFT collection in terms of terms of alpha value, p-value and test statistic of the Kolmogorov-Smirnov test. Values are provided considering separately both input and output degree distribution. The merged graph considers all collections together.}
\label{t:graph:powerlaw}
\resizebox{0.42\textwidth}{!}{
 
\begin{tabular}{|l|c|c|c|c|c|c|}\hline
\multirow{2}{*}{\textbf{Collection}} & \multicolumn{3}{c|}{\textbf{Input}}          & \multicolumn{3}{c|}{\textbf{Output}}         \\
                                     & \textbf{alpha} & \textbf{p} & \textbf{stat} & \textbf{alpha} & \textbf{p} & \textbf{stat} \\\hline
HashMasks	& 2.64& 	0.83& 	0.05& 	2.46& 	0.35& 	0.07\\\hline
Art Blocks Curated& 	3.18& 	0.98& 	0.05	& 3.37& 	0.97& 	0.05\\\hline
CryptoPunks	& 2.41& 	0.76& 	0.05& 	2.18& 	0.87& 	0.05\\\hline
Bored Ape Yacht Club & 	2.68 & 	0.86& 	0.04 & 	2.60 & 	0.82& 	0.06\\\hline
Acclimated MoonCats& 	2.39& 	0.77& 	0.05& 	2.30& 	0.60& 	0.09\\\hline
CryptoVoxels& 	2.47& 	0.81& 	0.05& 	2.30& 	0.96& 	0.05\\\hline
Decentraland& 	1.87& 	0.98& 	0.05& 	2.22& 	0.86& 	0.05\\\hline
Meebits	& 2.21& 	0.70& 	0.048& 	2.51& 	0.56& 	0.06\\\hline
ALL (merged)& 	2.79& 	0.97& 	0.03& 	2.72& 	0.99& 	0.028 \\\hline         
\end{tabular}
 
}
\end{center}
\end{table}

%% file: tables/graph_connectivity.tex
\begin{table}[htbp]
\begin{center}
\caption{Network graph connectivity and clustering statistics for each NFT collection in terms of terms of reciprocity, transitivity and assortativity. The merged graph considers all collections together.}
\label{t:graph:connectivity}
\resizebox{0.42\textwidth}{!}{
\begin{tabular}{|l|c|c|c|} \hline
\textbf{Collection} & \textbf{Reciprocity} & \textbf{Transitivity} & \textbf{Assortativity} \\\hline
HashMasks	 & 0.049	& 0.019 & -0.015\\\hline
Art Blocks Curated& 0.040	 & 0.039	 & 0.313\\\hline
CryptoPunks	& 0.044 & 0.058	 & 	0.474\\\hline
Bored Ape Yacht Club & 0.058	 & 0.012 	 & 	0.026 \\\hline
Acclimated Moon Cats& 0.021	 &  0.032	 & 0.616\\\hline
CryptoVoxels & 0.048	 & 0.012	 &  0.020\\\hline
Decentraland & 0.071	 & 0.037	 &  0.109\\\hline
Meebits &  0.064	 & 0.012	 & 	 0.057\\\hline
ALL (merged) & 0.054	 & 0.023	 &  0.045   \\\hline               
\end{tabular}
}
\end{center}
\end{table}

%% file: tables/soa_graphs_stats.tex
\begin{table}[htbp]
\begin{center}
\caption{Estimated power-law coefficient, diameter and mean distance of web, social networks and ERC20 token networks.}
\label{t:graph:soa}
\resizebox{0.4\textwidth}{!}{
\begin{tabular}{|l|c|c|c|c|} \hline
\multirow{2}{*}{\textbf{Network}}  & \multicolumn{2}{c|}{\textbf{alpha}}     & \multirow{2}{*}{\textbf{Diameter}} & \multirow{2}{*}{\textbf{Mean Distance}} \\
                                   & \textbf{Outdegree} & \textbf{Indegree} &                         &  \\ \hline
                                  
Web~\cite{web_graph_andrei}                         & 2.09 & 2.67 & 905.00 & 16.2   \\ \hline  
Social networks~\cite{social_network_graph_mislove} & 1.80 & 1.63 & 22.66  & 5.55   \\ \hline     
ERC-20 tokens~\cite{eth_graph_chen}                 & 2.80 & 2.60 & -      & 2          \\ \hline     
\end{tabular}
}
\end{center}
\end{table}

%% file: tex/time_behaviour.tex
\section{Time behaviour analysis}
\label{s:timeBehaviour}
This section discusses some results obtained by analysing NFT transaction graphs and taking into account the timestamps at which transactions are made. In particular, a methodology is defined to: estimate how both the value of an entire collection and of a single wallet vary with time; calculate the daily volume exchanged to identify the major owners of tokens in a collection; follow token accumulation and sale flows.

\begin{figure*}
\begin{center}
\subfloat[Collection value\label{f:time:results:collValue}]{\includegraphics[width = 0.23\linewidth]{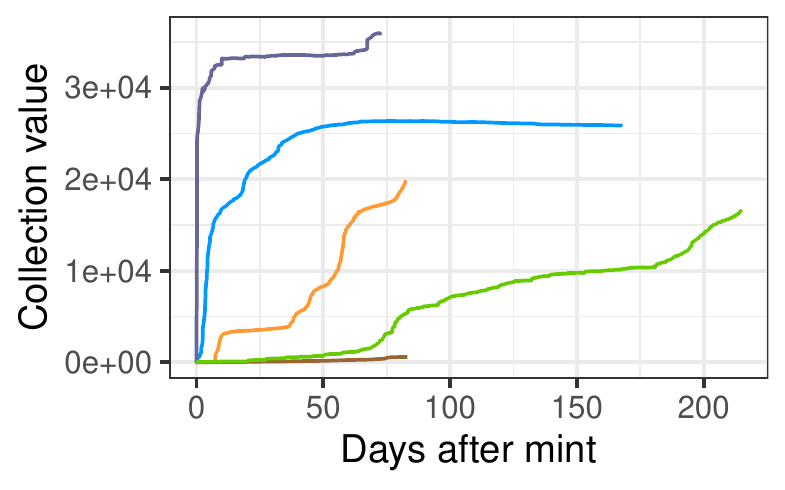}} \quad  
\subfloat[Average wallet value\label{f:time:results:walletValue}]{\includegraphics[width = 0.23\linewidth]{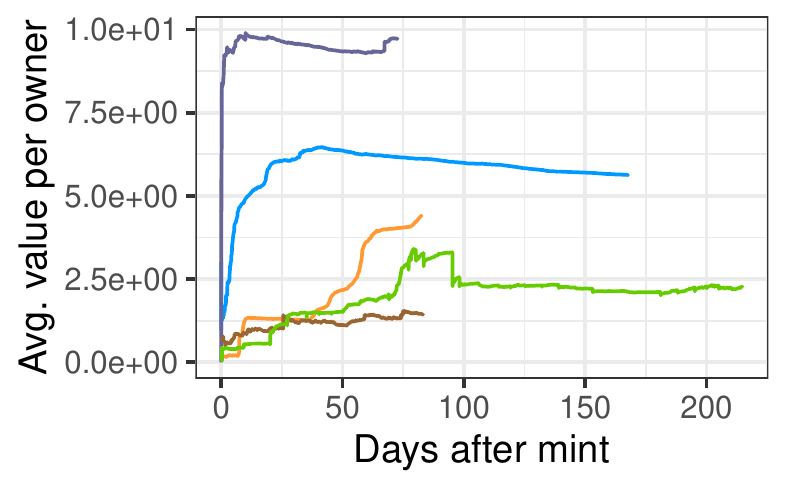}}\quad 
\subfloat[Unique wallets\label{f:time:results:uniqueOwners}]{\includegraphics[width = 0.23\linewidth]{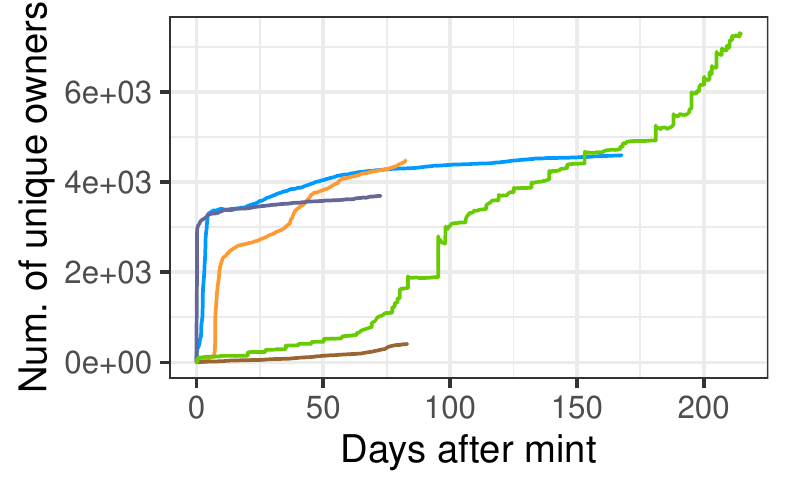}} \quad  
\subfloat[Average tokens per wallet\label{f:time:results:tokensWallet}]{\includegraphics[width = 0.23\linewidth]{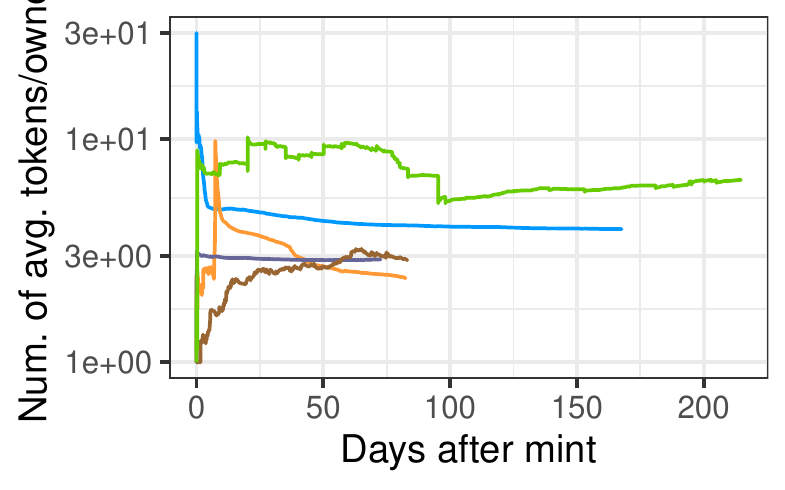}} \\
\subfloat[Daily transactions \label{f:time:results:trx}]{\includegraphics[width = 0.23\linewidth]{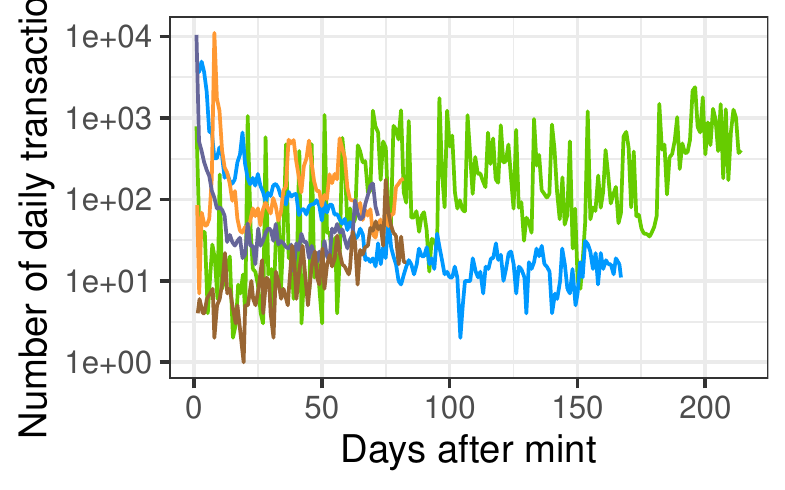}} \quad
\subfloat[Daily volume (ETH)\label{f:time:results:volume}]{\includegraphics[width = 0.23\linewidth]{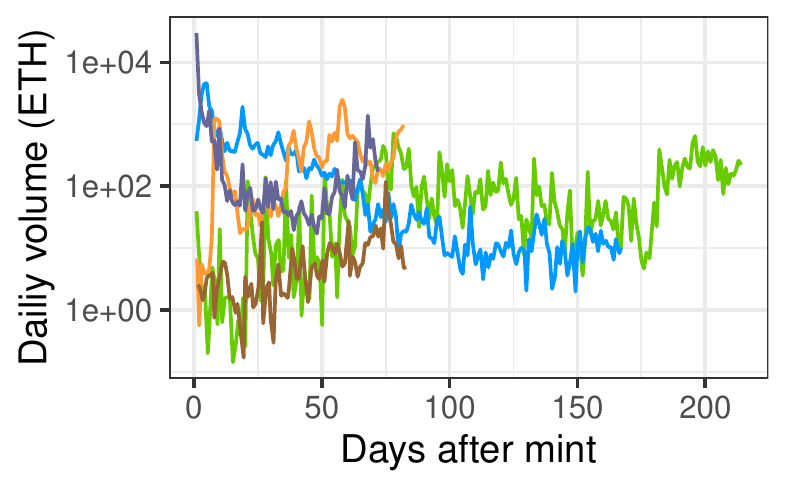}} \quad  
\subfloat[Average price multiplier with respect to mint\label{f:time:results:multiFromMint}]{\includegraphics[width = 0.23\linewidth]{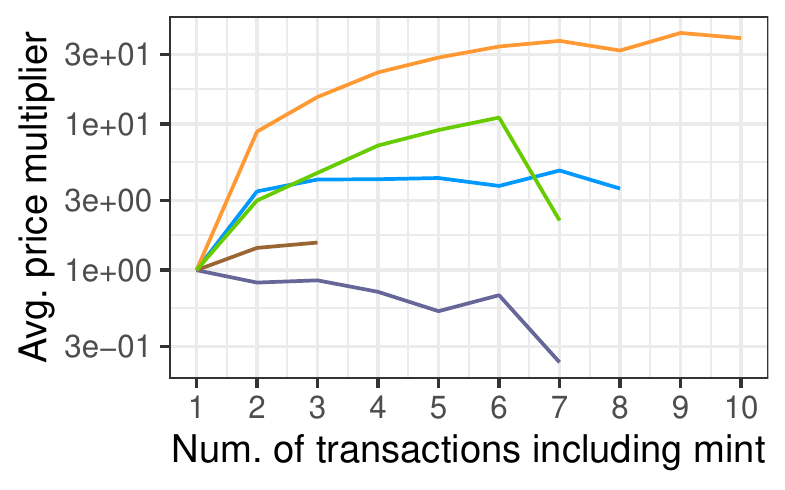}} \quad  
\subfloat[Average price multiplier with respect to fisrt transaction\label{f:time:results:multiFromFirstTx}]{\includegraphics[width = 0.23\linewidth]{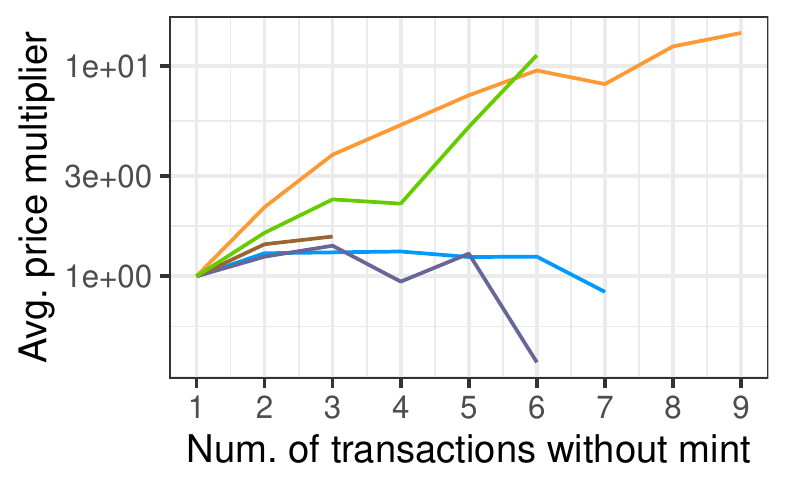}} \\
\subfloat[Number of transactions per token distribution, including mint\label{f:time:results:tokensTransaction}]{\includegraphics[width = 0.23\linewidth]{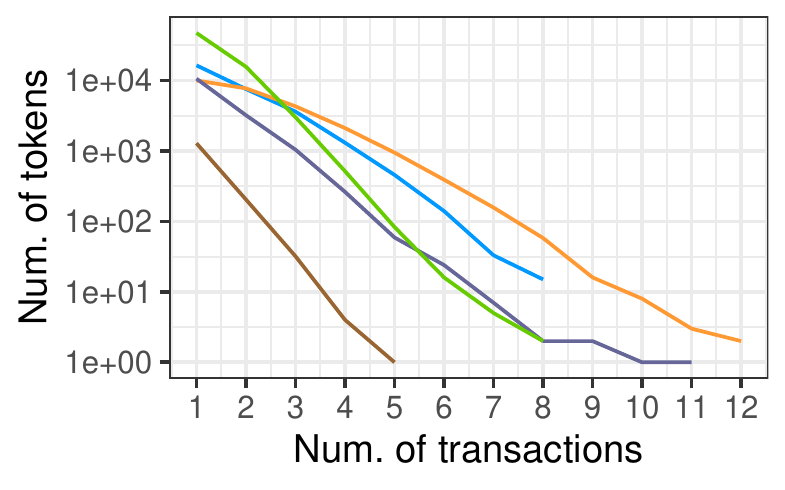}} \quad  
\subfloat[HashMasks 5 top owners flow\label{f:time:results:ownHM}]{\includegraphics[width = 0.23\linewidth]{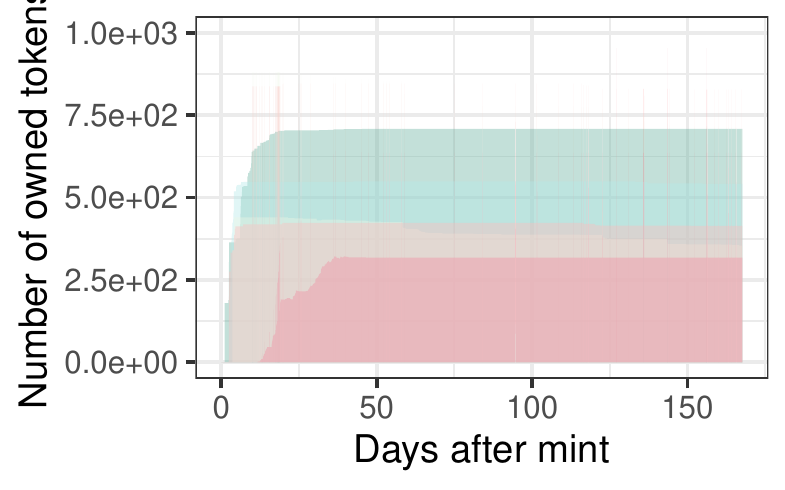}}  \quad  
\subfloat[Bored Apes 5 top owners flow\label{f:time:results:ownBAYC}]{\includegraphics[width = 0.23\linewidth]{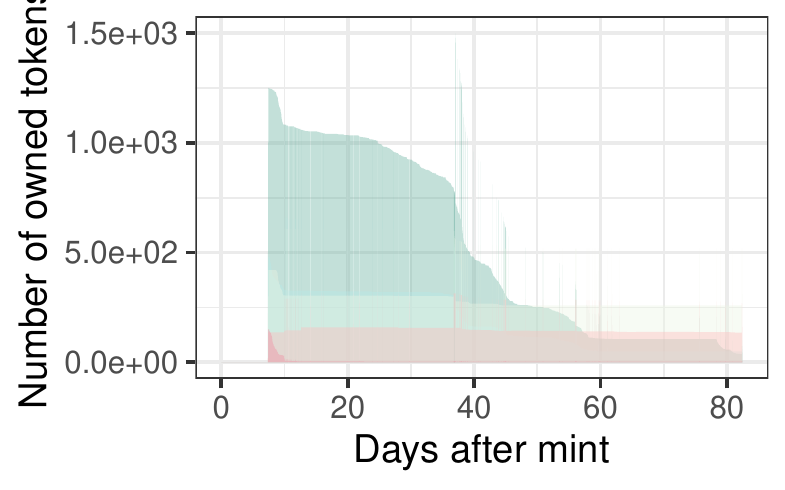}} \quad  
\subfloat[Meebits 5 top owners flow\label{f:time:results:ownMeebits}]{\includegraphics[width = 0.23\linewidth]{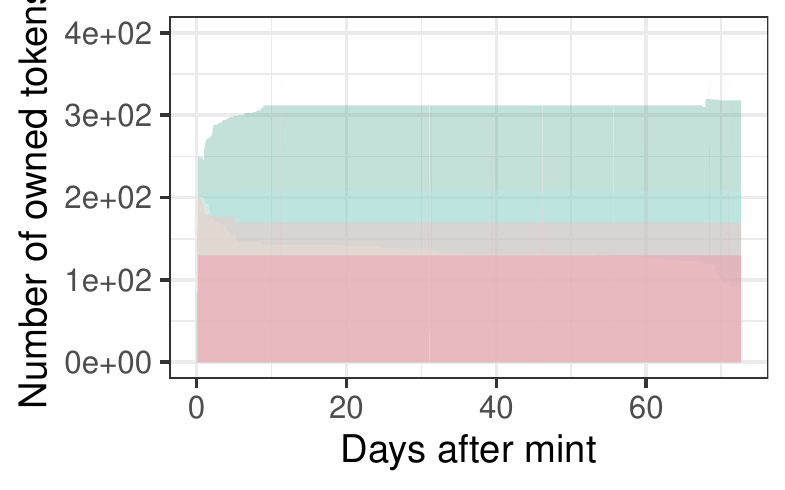}}
\caption{Time behaviour analysis for HashMasks, Bored Apes, Meebits, MoonCats, and Artblocks. The colours are the same as those of the legend in Figure~\ref{f:graph:degree}.}
\label{f:time:results}
\end{center}
\end{figure*}

\subsection{Collection and wallets value}
As with other types of assets such as cryptocurrencies or stocks, one of the most relevant quantities when analyzing an NFT project is how its value evolves over time. In the case of ERC721 tokens, the valuation can be done on several levels: (1) analyzing price evolution for each single token; (2) analyzing the evolution of the value of the whole collection. The latter can be calculated as the sum of all the last transaction price of the tokens in a collection. Self-transactions and zero-cost transactions (previously also referred to as transfers) are excluded from the calculation. Figure~\ref{f:time:results:collValue} illustrates the value trend for selected projects from Table~\ref{t:nft:names} (HashMasks, Bored Apes, Meebits, MoonCats, and ArtBlocks). This analysis can then be applied to the individual wallets that have at least one token of a given collection: results are illustrated in Figure~\ref{f:time:results:walletValue}. They show the widespread presence across a number of NFT networks  of both investors accumulating NFTs and individuals who make large profits. Another important metric for a project is the time evolution of the number of unique wallets holding at least one token from the collection -- for the NFT projects under consideration, this is shown in Figure~\ref{f:time:results:uniqueOwners}. As for typical successful projects, both this number and the overall value of the network grow with time.

\subsection{Exchanged volume and number of transactions}
In order to understand how interesting a collection is, especially during hype cycles, one of the fundamental values that must be observed (in addition to the number of unique owners) is the number of daily transactions together with the daily volume traded. These values are calculated by summing up the costs of transactions that have been executed on the same day. For the collections examined, these two trends are shown in Figure~\ref{f:time:results:trx} and Figure~\ref{f:time:results:volume}, respectively.
By analysing the links in the graph, it is also possible to estimate two other parameters that indicate the evolution of the value of a collection: the average price multiplier for a token with respect to the cost of minting or exchanging it for the first time, and the average number of transactions (including minting) for each token in the collection. The results are shown in Figure~\ref{f:time:results:multiFromMint} and Figure~\ref{f:time:results:multiFromFirstTx}. As can be seen, the earlier a purchase is made (i.e., during the mint or just after), the greater the probability of making a profit by reselling a token. The results of the second analysis are summarized in Figure~\ref{f:time:results:tokensTransaction}: in general, a large proportion of the tokens in each collection were bought at minting time and have not been sold since. This is certainly an indicator of how much NFT technology is in its infancy. An alternative explanation is that NFTs are seen as a store-of-value investment, both to be resold only at some future time. Interestingly though, looking at the example of the Bored Ape Yacht Club NFTs, which have the highest exchange probability, they are also characterized by a substantial increase in the value of the collection in a very short time frame (see Figure~\ref{f:time:results:collValue}).

\subsection{Flows of major owners}
One of the big differences between ERC-20 (fungible) and ERC-721 (non-fungible) tokens is that the latter are an asset generally made for display, as many of these collections have very high entry prices (see e.g. the Cryptopunks floor price from Table~\ref{t:nft:collected_data}). This is why even different NFTs are typically accumulated in the same wallet (e.g., to be displayed in online galleries and associated to a social media profile). It is therefore possible to identify for a given collection which wallets have had the largest amount of tokens over time since the minting date, and follow the flow of sales and purchases.  Figures~\ref{f:time:results:ownHM},~\ref{f:time:results:ownBAYC} and~\ref{f:time:results:ownMeebits} illustrate the evolution of the 5 top wallets for the collections of HashMasks, Bored Ape Yacht Club, and Meebits respectively. HashmMask and Meebits have a very similar trend, where the first 3 wallets have consolidated their position during the time; on the other hand, for the Bored Ape Yacht Club the top wallet~\cite{Pranksy_account} has, after an initial dormant phase, put all its tokens on the market, making the value of the whole collection increase as a consequence (Figure~\ref{f:time:results:collValue}). This wallet is one of a number of successful investors who did succeed at extracting large profits from NFT collections.

%% file: tex/conclusions.tex
\section{Conclusions}
\label{s:conclusion}
In this work, the first systematic analysis of transactions in an NFT ecosystem based on the Ethereum blockchain has been presented. Some of today's best known NFT projects (such as CryptoPunks, HashMasks and ArtBlocks) have been analysed. 
By filtering information publicly available on the blockchain, and structuring it in a graph model, one can perform systematic analysis and define metrics that help establishing how the wallets participating in this ecosystem interact. It was shown that these graphs follow a power-law in their nodes degree distribution. Topological values such as graph diameter and mean distance were then compared with state of art results for graphs derived for web networks, social networks and ERC20 token network. That led to the conclusion that the structure of NFT networks is qualitatively very similar to the one measured for interactions in social networks. Furthermore, a methodology was provided to identify the major NFT owners and follow their buying and selling flows. In fact, this study can help shed quantitative light on a market which might otherwise be prone to hype and misleading information.

Further research may refine our understanding of wallets holding many NFTs. In this work we were able to identify them, but further investigation is needed to identify how their strategies (in terms of accumulation and subsequent selling) on one or more collections affect the markets and the overall value of the different NFTs. In addition, the analysis might be refined by considering the total costs of each transaction (e.g., split payments when the transaction goes through a smart contract, and Ethereum gas fees). However, doing so will not change the main conclusions of this study.